\documentclass[epj]{svjour}
\usepackage{latexsym}
\usepackage{graphics}
\usepackage{amsmath,amssymb}
\usepackage{graphicx}
\usepackage{dcolumn}
\usepackage{bm}
\usepackage{psfrag}
\usepackage{subfigure}
\usepackage{epstopdf}
\usepackage{mathbbol}
\newcommand{\Tr}{\text{Tr}}
\usepackage{color}
\definecolor{rot}{rgb}{0.75,0.05,0.25}
\definecolor{hellgrau}{gray}{0.5}
\definecolor{blau}{rgb}{0,0,0.7}
\definecolor{LightCyan}{rgb}{0.88,1,1}

\begin{document}
\title{Quantum Hertz entropy increase in a quenched spin chain}
\author{Darshan G. Joshi\inst{1} \and Michele Campisi\inst{2}
}                     
%
%
\institute{Institut f\"ur Theoretische Physik, Technische Universit\"at Dresden, D-01062 Dresden, Germany \and Institute of Physics, University of Augsburg, D-86153 Augsburg, Germany}
\date{Received: date / Revised version: date}
%
\abstract{
The classical Hertz entropy is the logarithm of the
volume of phase space bounded by the constant energy surface; its quantum
counterpart, the quantum Hertz entropy, is $\hat S = k_B \ln \hat N$, 
where the quantum operator $\hat N$ specifies the 
number of states with energy below a given energy eigenstate.
It has been recently proved that, when an isolated quantum mechanical system is driven out of equilibrium 
by an external driving, the change in the expectation of its quantum Hertz 
entropy cannot be negative, and is null for adiabatic driving. This is in full 
agreement with the Clausius principle.
Here we test the behavior of the expectation
of the quantum Hertz entropy in the case when two identical $XY$ spin chains initially
at different temperatures are quenched into a single $XY$ chain. 
We observed no quantum Hertz entropy decrease.
This finding further supports the statement that the quantum Hertz entropy
is a proper entropy for isolated quantum systems.
We further quantify how far the quenched chain is from thermal equilibrium
and the temperature of the closest equilibrium.
} 
\maketitle

%

\section{Introduction}
The recent tremendous development in the field of nonequilibrium quantum fluctuations
\cite{Campisi11RMP83,Esposito09RMP81}, has unveiled with an unprecedented clarity 
that many phenomena traditionally associated exclusively with macroscopic 
thermodynamic behaviour may manifest themselves even at
the microscopic quantum level. Notably, the Second Law of thermodynamics, in the 
work-free energy formulation \cite{Fermi56Book},
\begin{equation}
\langle w \rangle \geq \Delta F \, ,
\label{eq:kelvin}
\end{equation}
holds down to the quantum level \cite{Allahverdyan02PHYSA305}. In Eq. (\ref{eq:kelvin}), $w$ is the work 
done on a quantum system that is initially in equilibrium with a thermal bath, when the system
is perturbed by an external time dependent protocol that changes its Hamiltonian in time.
The brackets indicate average over many realizations, and $\Delta F$ is the 
difference between the free energy of a hypothetical equilibrium state
(not necessarily reached by the system) corresponding to the final Hamiltonian, 
and the actual initial free energy of the system.
Eq. (\ref{eq:kelvin}) follows straightforwardly from the quantum version
of the Jarzynski identity \cite{Jarzynski97PRL78,Tasaki00arXiv,Kurchan00arXiv,Talkner07JPA40}.

For a cyclical driving, $H(\tau)=H(0)$, $\Delta F=0$,
Eq. (\ref{eq:kelvin}) says, in accordance with Kelvin postulate, that no energy
can be extracted by the cyclic variation of a parameter from a system
in contact with a single bath \cite{Allahverdyan02PHYSA305}:
\begin{equation}
\langle w \rangle \geq 0 \quad \text{Kelvin (cyclic, with bath)}
\label{eq:clausius}
\end{equation}
Given the recent theoretical and experimental
advances concerning the nonequilibrium dynamics of isolated quantum systems
\cite{Polkovnikov11RMP83}, an interesting question is whether 
a microscopic quantum formulation of the second law in accordance 
with Clausius formulation, is possible as well. According to 
Clausius' formulation the change of entropy of a \emph{thermally
insulated} driven system, which begins and ends in equilibrium,
is non-negative:
\begin{equation}
\Delta S \geq 0 \quad \text{Clausius (no bath)}
\label{eq:clausius}
\end{equation}
Answering this question is not a simple task because it amounts to
singling out a quantum mechanical quantity that behaves as prescribed
by the Clausius principle and goes over to the usual thermodynamic entropy 
in the classical/thermodynamic limit. Of course von Neumann
``entropy'', $-\Tr \rho \log \rho$, proves inadequate in this respect because it is invariant 
under the quantum unitary time evolution.

One proposal in the direction of answering the above question
was put forward by Polkovnikov \cite{Polkovnikov11AP326,Santos11PRL107},
with the introduction of the so-called diagonal entropy.
Another proposal, put forward by Hal Tasaki and by one of us \cite{Tasaki00arXivb,Campisi08SHPMP39,Campisi08PRE78b},
uses instead the Hertz entropy
 \cite{Hertz10AP338a,Campisi05SHPMP36,Campisi10AJP78}
see Eq. (\ref{eq:Hertz}), or in quantum mechanics,
its quantum counterpart, that is the logarithm of the ``quantum number operator''  $\hat{N}$ \cite{Tasaki00arXivb,Campisi08SHPMP39,Campisi08PRE78b},
see Eq. (\ref{eq:hatS}) below.
In \cite{Tasaki00arXivb,Campisi08SHPMP39,Campisi08PRE78b} it was shown that the
expectation of the quantum Hertz entropy 
behaves in accordance to the Clausius principle.
Here we scrutinize whether it complies also with another 
property of thermodynamic entropy, namely whether it increases
in a scenario when 
two quantum systems initially at different temperatures are allowed to exchange energy
via the sudden switch-on of an interaction.
This is a scenario that has recently attracted considerable attention
\cite{Ponomarev11PRL106,Ponomarev12EPL98}, and that, given the recent advances,
e.g., in ultra-cold-atom physics \cite{Bloch08RMP80},
is amenable to experimental investigations.

In our study the two interacting bodies are two
isotropic $XY$ spin chains of length $N/2$, which are initially 
at different temperatures, and are
suddenly quenched into a single isotropic $XY$ spin chain of length $N$.
While the quench dynamics in spin-chains has been thoroughly studied,
only few studies addressed their thermodynamics 
\cite{Campisi10CP375,Dorner12PRL109}.

In Secs. \ref{sec:quantum-entropy} and  \ref{sec:model} we review the quantum Hertz
entropy and present our model, respectively.
In Secs. \ref{sec:InitialEntropy} and \ref{sec:FinalEntropy} we calculate
the initial and final expectation of the quantum Hertz entropy. 
Results concerning the entropy change and the deviation of the
final state from thermal equilibrium are presented in Sec. \ref{sec:entropychange}
and Sec. \ref{sec:thermalization}, respectively. Conclusions are drawn in
Sec. \ref{sec:conclusions}.

\section{Quantum Hertz entropy}\label{sec:quantum-entropy}
The microcanonical entropy of a classical system is the so-called Hertz entropy 
\cite{Hertz10AP338a,Campisi05SHPMP36,Campisi10AJP78,Hilbert06PRE74}
presented also by Gibbs in his classic book \cite{GibbsBook},
namely 
\begin{align}
S= k_B \ln [\Phi/h^{f}] \, ,
\label{eq:Hertz}
\end{align}
 where $k_B$ is Boltzmann constant, $\Phi$ is the phase space volume enclosed by the 
hyper-surface of constant energy in the system phase space of dimension $2f$ and $h$ is Planck's constant. 
According to semiclassical reasoning \cite{LandauBook5}, in the quantum limit, 
the quantity $\Phi / h^{f}$ gets the discrete values $n=1,2,\dots$ .\footnote{Depending
on the problem at hand, the quantization rule may
prescribes a shift $\Phi / \hbar \rightarrow n+a$, which is not relevant in this context.} 
The associated quantum operator is the quantum number operator $\hat N$, whose
eigenvectors are the energy eigenvectors, with the integers the corresponding eigenvalues.
For a driven system, the  operator $\hat N$ is
time dependent, and its spectral decomposition reads (for non degenerate Hamiltonians)
$\hat N(t)= \sum n\Pi_n(t)$, with $\Pi_n(t)$ the instantaneous eigenprojectors on the
eigenspace spanned by the instantaneous eigenvalue $E_n(t)$ of the Hamiltonian $H(t)$.
Here it is assumed that the energy eigenvalues are non-degenerate and ordered in increasing fashion:
$E_1(t)<E_2(t)<\dots $ . 
Accordingly, the quantum mechanical operator associated to the Hertz entropy is 
\begin{eqnarray}
\hat S(t) &= k_B \ln \hat N(t)
 \label{eq:hatS} \, .
\end{eqnarray}

Under the assumption that (i) the density matrix $\rho(0)$ describing the initial 
state of the system is diagonal in the energy eigenbasis $\rho(0)= \sum p_n\Pi_n(0)$,
and (ii) the population $p_n$ decreases with increasing energy (i.e., $p_n \leq p_m$, if $E_n(0) > E_m(0) $);
it has been shown that \cite{Tasaki00arXivb,Campisi08SHPMP39,Campisi08PRE78b}:
\begin{equation}
S(t) \geq S(0) \, ,
\label{eq:clausius}
\end{equation}
where 
\begin{eqnarray}
S(t) &=  \Tr \hat S(t) \rho(t) \label{eq:S} \, .
\end{eqnarray}
For adiabatic transformations the equal sign holds in Eq. (\ref{eq:clausius}).
The quantity $S(t)$ hence behaves in accordance to the Clausius
principle and goes over to the usual thermodynamic entropy 
in case of large classical systems at equilibrium. 
These facts make it a sound
quantum mechanical counterpart for the thermodynamic entropy of a thermally insulated system.
In the following we shall refer to $S$ and $\hat S$ as to the quantum entropy, and
the quantum entropy operator, respectively. Note however that unless the system is 
in equilibrium at time $t$, the quantum entropy $S(t)$ should not be considered as
the thermodynamic entropy of the system. The latter is an exclusively equilibrium 
property.

\subsection{Remarks}
We remark that unlike the Boltzmann entropy
$S_B = k_B$ $\log [\Omega/h^{f}]$,
[$\Omega = \partial_E \Phi$ being the density of states], the Hertz entropy \emph{is not
postulated}, but rather \emph{rationally derived} from the fundamental requirement that its differential $dS$
exactly equals the quantity $\delta Q /T$ as calculated in the microcanonical ensemble, the so-called generalized Helmholtz 
theorem \cite{Campisi05SHPMP36,Campisi10AJP78}. As such, at equilibrium, the Hertz entropy has to be identified with the 
thermodynamic entropy. It is commonly assumed that $S$ and $S_B$ are equivalent \cite{Huang87Book},
which is true in most cases but has important exceptions, e.g., in small systems or systems with a finite spectrum.
In this latter case they can give drastically different results, notably $S$, a monotonically increasing function of $E$, 
gives only positive temperatures in accordance to thermodynamic fundamentals \cite{Callen60Book},
whereas $S_B$ predicts also negative ones. This is a topic of current interest as testified by recent experiments \cite{Braun13SCIENCE339}.

For a Gibbs ensemble of systems at canonical temperature $T_c= 1/k_B \beta$, one finds the 
following expression for the work dissipated, $W_\text{diss} = \langle w \rangle -\Delta F$, due to
an external driving \cite{Schloegl66ZP191,Bochkov81aPHYSA106,Kawai07PRL98,Deffner10PRL105}:
\begin{align}
 W_\text{diss} &= k_B T_c D[\rho(t)|\rho^\text{eq}(t)]
 \label{eq:Wdiss}
\end{align}
where $D[\rho(t)|\rho^\text{eq}(t)] = \Tr \rho(t) \log \rho(t) - \Tr \rho(t) \log \rho^\text{eq}(t) $
is the Kullbeck-Leibler divergence between $\rho(t)$, the density matrix of the system at 
time $t$, and $\rho^\text{eq}(t)=e^{-\beta H(t)}/\Tr e^{-\beta H(t)}$ the corresponding equilibrium density matrix.
The quantity $W_\text{diss}/T_c=k_B D[\rho(t)|\rho^\text{eq}(t)]$ is often referred to as the entropy production 
\cite{Crooks99PRE60}. This off-equilibrium quantity should not be confused with the change in thermodynamic entropy $\int \delta Q/T$, which involves quasi-static heat exchanges, instead.
However there exist a strict connection between $W_\text{diss}$ and the change in thermodynamic entropy $\Delta S$.

Noticing that the Kullbeck-Leibler divergence is a non-negative quantity, it is apparent from Eq. (\ref{eq:Wdiss}) that
$W_\text{diss} \geq 0$ for $T_c> 0$, $W_\text{diss} \leq 0$ for $T_c< 0$.
This can be easily obtained also from the Jarzynski identity \cite{Campisi11RMP83}.
It says that when one perturbs a Gibbs ensemble  of thermally isolated systems characterized by a positive canonical $T_c$,
one gets out less energy than one puts in, on average.
This is the second law in the usual Kelvin formulation.
Vice-versa one gets out more energy than one puts in, if the ensemble was characterized by a 
negative $T_c$.\footnote{Note that ``negative temperature can never exist in equilibrium states. It is however,
possible to create it in certain transient processes" \cite[see pag. 148]{Kubo65book}.} This is well described by Ramsey's account 
of the early experiments on negative spin temperatures. He writes \cite[see p. 27]{Ramsey56PR103}  
``when a negative temperature spin system was subjected to resonance radiation 
more radiant energy was given off by the spin system than was absorbed.''
In sum, it is an experimental fact that when perturbing canonical ensembles characterized by negative $T_c$, the second law is inverted. 
Accordingly the entropy should decrease, and this is in fact the behavior of the
Hertz entropy \cite{Campisi08SHPMP39}:
$\Delta S \geq 0$ for $T_c> 0$, $\Delta S \leq 0$ for $T_c< 0$,
which shows the equivalence of Kelvin and Clausius formulations down to microscopic
quantum level, and also to negative $T_c$'s.

One difference between the quantum Hertz entropy and
the diagonal entropy \cite{Polkovnikov11AP326} is that the latter would always increase, regardless of
the sign of $T_c$ of the initial canonical ensemble. Presumably the diagonal entropy does not differ much from the Hertz entropy
in ordinary large systems with unbound spectrum. This question however deserves
a detailed investigation that goes beyond the scope of the present contribution.

\section{The Model}\label{sec:model}
\begin{figure}[b]
	\centering
		\includegraphics[width=.48\textwidth]{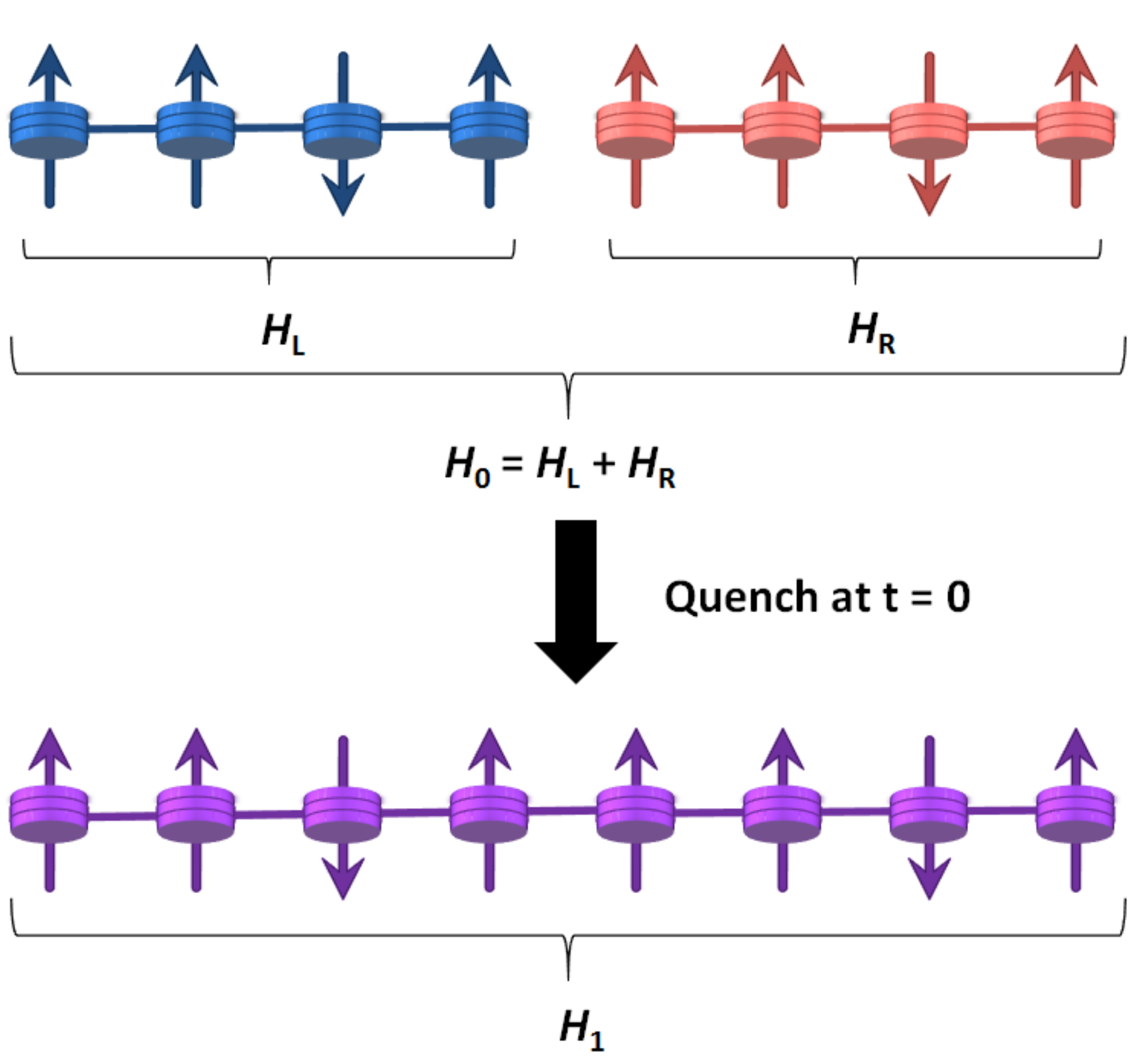}
	\caption{Schematics of our model}
	\label{fig:chain}
\end{figure}

Our model consists of two identical isotropic XY spin chains, which are
quenched at time $t=0$ to a single XY chain of twice the length,
see Fig. \ref{fig:chain}.
At times $t<0$, the system Hamiltonian is:
\begin{equation}
H_0 = H_L + H_R \, ,
\end{equation}
with:
\begin{align}
H_L &= -\frac{h}{2}\sum_{j=1}^{N/2} {\sigma_j ^z} + \frac{J}{2}\sum_{j=1} ^{N/2-1} {[\sigma_j ^x \sigma_{j+1} ^x + \sigma_j ^y \sigma_{j+1} ^y]}\, ,\\
H_R &= -\frac{h}{2}\sum_{j=N/2+1}^{N} {\sigma_j ^z} + \frac{J}{2}\sum_{j=N/2+1} ^{N-1} {[\sigma_j ^x \sigma_{j+1} ^x + \sigma_j ^y \sigma_{j+1} ^y]}\, .
\end{align}
Here $\sigma_j^\alpha$, $j=1 \dots N$, $\alpha=x,y,z$, denotes the Pauli matrices of  the $j$-th spin.
At $t<0$ the left (right) chain is in the Gibbs state of  temperature $T_{L(R)}$, hence the total system density matrix is
given by their product:
\begin{equation}
\rho_0 = \frac{e^{-\beta_L H_L}}{Z(\beta_L)} \otimes \frac{e^{-\beta_R H_R}}{Z(\beta_R)} \, ,
\label{eq:initialstate}
\end{equation}
where $Z(\beta) = \Tr e^{-\beta H_{L} }= \Tr e^{-\beta H_R }$ is the partition function,
and $\beta_{L(R)}=(k_BT_{L(R)})^{-1}$, with $k_B$ Boltzmann constant.
At time $t=0$ an interaction between spin $N/2$ and spin $N/2+1$ is turned on,
such that the Hamiltonian is, for $t>0$:
\begin{align}
H_1 &= -\frac{h}{2}\sum_{j=1}^{N} {\sigma_j ^z} + \frac{J}{2}\sum_{j=1} ^{N-1} {[\sigma_j ^x \sigma_{j+1} ^x + \sigma_j ^y \sigma_{j+1} ^y]}\, .
\end{align}

The Hamiltonians $H_L,H_R,H_1$ all represent isotropic XY chains of different lengths.
Following the standard procedures they can be put in diagonal form by 
means of Jordan-Wigner transformation followed by a sine transform \cite{Lieb61AP16,Mikeska77ZPB26}.
Specifically
\begin{equation}
H_1 =  \sum_{j=1}^{N}\varepsilon_k  b_k^{\dagger}b_k + \frac{N h}{2} \, ,
\label{eq:H_1}
\end{equation}
where 
\begin{align}
\varepsilon_k &= - h - 2J \cos \left( \frac{k\pi}{N+1} \right)\, ,\\
 b_k &= \sqrt{\frac{2}{N+1}} \sum_{i=1}^{N} \sin \left( \frac{ki\pi}{N+1} \right)a_i\, , \\
 a_j &= \prod _{k=1}^{j-1}\sigma^z_k \sigma_j^{-} \, ,\\
 \sigma_j^{\pm}&= \frac{1}{2}(\sigma^x_j\pm i\sigma^y_j) \, ; \label{eq:sigmas}
\end{align}
We shall denote the eigenvectors and eigenvalues of $H_1$ as $ |\bm n  \rangle= |n_1, n_2, \dots n_N  \rangle$ and
$E_{\bm n}$, respectively, where $n_i=0,1$:
\begin{align}
H_1 |\bm n  \rangle = E_{\bm n}  |\bm n  \rangle, \qquad 
E_{\bm n} = \sum_{k=1}^{N}\varepsilon_k  n_k + \frac{N h}{2}\, .
\end{align}
The states $ |\bm n  \rangle$ are the Fock states associated to the
fermonic operators $b_k$. For future reference we recall their properties
\begin{align} 
b_k^\dagger b_k  |...\,  n_k \,... \rangle &= n_k | ...\,  n_k \, ...  \rangle \label{eq:bk1} \\
b_k |...\,  n_k \,...  \rangle &= n_k (-1)^{\sum_{i=1}^{k-1}n_i} | ...\, n_k-1 \,...  \rangle  \label{eq:bk2}\\
b_k^\dagger |...\,  n_k \,...  \rangle &= (1-n_k)(-1)^{\sum_{i=1}^{k-1}n_i} | ...\,  n_k+1 \,...  \rangle  \label{eq:bk3}
\end{align}

Similarly, for the $L$-chain:
\begin{align}
H_L &=  \sum_{k=1}^{N/2}\varepsilon'_k  b_k'^{\dagger}b'_k + \frac{N h}{4} \, ,
\label{eq:H_L}\\
\varepsilon_k' &= - h - 2J \cos \left( \frac{k\pi}{N/2+1} \right)\, , \\
b'_k &= \sqrt{\frac{2}{N/2+1}} \sum_{i=1}^{N/2} \sin \left( \frac{ki\pi}{N/2+1} \right)a_i \, .
\end{align}
Note that the same Jordan-Wigner operators $a_i$ are used for the $L$-chain and the total chain.
In defining $b_k$ all $N$ operators are used, while only the first $N/2$ are employed to define the primed
operators $b'_k$.
We shall denote the eigenvectors and eigenvalues of $H_L$ as $ |\bm l  \rangle= |l_1, l_2, \dots l_{N/2}  \rangle$ and
$E_{\bm l}$, respectively:
\begin{align}
H_L |\bm l  \rangle = E_{\bm l}  |\bm l  \rangle, \qquad 
E_{\bm l} = \sum_{k=1}^{N/2}\varepsilon'_k  l_k + \frac{N h}{4}\, .
\end{align}
Likewise for the $R$-chain, 
\begin{align}
H_R &=  \sum_{k=1}^{N/2}\varepsilon'_k 
b_k''^{\dagger}b''_k + \frac{N h}{4} \, ,
\label{eq:H_L}\\
b''_k &= \sqrt{\frac{2}{N/2+1}} \sum_{i=1}^{N/2} \sin \left( \frac{ki\pi}{N/2+1}
\right) a''_i\\
a''_j &= \prod _{k=1}^{j-1}\sigma^z_{N/2+k} \sigma_{N/2+j}^{-}
\end{align}
Note two prominent facts: i) the single mode eigenenergies $\varepsilon'_k$ are the same for the $L$-chain and the $R$-chain, because the two chains are identical.
(ii) The Jordan Wigner operators $a''_j$ of the $R$-chain differ from the Jordan Wigner operators $a_j$
of the $L$-chain and total chain, because the latter begin with spin $1$, while the $R$-chain begins with spin $N/2+1$.
We shall denote the eigenvectors and eigenvalues of $H_R$ as $ |\bm r  \rangle= |r_1, r_2, \dots r_{N/2}  \rangle$ and
$E_{\bm r}$, respectively:
\begin{align}
H_R |\bm r  \rangle = E_{\bm r}  |\bm r  \rangle, \qquad
E_{\bm r} = \sum_{k=1}^{N/2}\varepsilon'_k  r_k + \frac{N h}{4}\, .
\end{align}
\section{Initial quantum Entropy} \label{sec:InitialEntropy}
At times $t<0$ the entropy of the two non-interacting chains is given by the sum
of their individual entropies. 
We proceed by calculating the quantum entropy $S^{\beta}$ of the left chain 
at inverse thermal energy $\beta$. Since the two chains are identical this also gives
the quantum entropy of right chain at the same inverse thermal energy $\beta$.

According to Eq. (\ref{eq:S}) 
\begin{align}
S^\beta &=  \frac{k_B}{Z(\beta)}\Tr \, e^{-\beta H_L}\, \ln \hat L \nonumber \\
&=  \frac{k_B}{Z(\beta)}\sum_{\bm l} \, e^{-\beta E_{\bm l}}\, \langle \bm l | \ln \hat L |\bm l \rangle \nonumber \\
&=  \frac{k_B}{Z(\beta)}\sum_{\bm l} \,  e^{-\beta E_{\bm l}} \, \ln \lambda_{\bm l} \, .
\label{eq:S[Kbeta]}
\end{align}
Here $\hat L$ is the principal quantum number operator associated to the $L$-chain.
Its eigenvectors are the Fock states $|\bm l \rangle $, and its eigenvalues are $\lambda_{\bm l}$.
The eigenvalues $\lambda_{\bm l}$ are calculated in the following way. 
The energy eigenvalues $E_{\bm l}$ are ordered
accordingly to their increasing values, so as to obtain a sequence
\begin{align}
E_{\bm l_1} < E_{\bm l_2}< \dots < E_{\bm l_{2^{N/2}}}\, ,
\end{align}
where $|\bm l_1\rangle $ is the ground state, $|\bm l_2\rangle $ is the first excited state, $\dots$ $|\bm l_{2^{N/2}}\rangle $
is the state of highest energy.
Then $\lambda_{\bm l_1} =1, \lambda_{\bm l_2} =2$, etc. 

The total initial quantum entropy $S_0$, is given by:
\begin{equation}
S_0 = S^{\beta_L}+ S^{\beta_R}.
\end{equation}

\section{Final quantum entropy}
\label{sec:FinalEntropy}
Due to the assumption of sudden quench, at time $t=0^+$ the density matrix
retains the initial form in Eq. (\ref{eq:initialstate}). The final quantum entropy $S_1$ is therefore given by
\begin{align}
S_1 &= k_B \frac{\Tr\, e^{-\beta_L H_L}  e^{-\beta_R H_R} \ln \hat N }
{Z(\beta_L)Z(\beta_R)} \nonumber \\
&= k_B \frac{\sum_{\bm n} \langle \bm n| e^{-\beta_L H_L}  e^{-\beta_R H_R} |\bm n\rangle \ln \nu_{\bm n} }
{Z(\beta_L)Z(\beta_R)}\, .
\end{align}
Here $\hat N$ is the quantum number operator associated to the total chain, and
$\nu_{\bm n}$ are its integer eigenvalues, obtained as described above for 
the smaller $L$-chain, with the difference that now one has to order the
$2^N$ eigenvalues $E_{\bm n}$.

We next consider the basis $|\bm l \bm r\rangle = |\bm l \rangle \otimes |\bm r\rangle$ formed by the direct
product of the eigenbasis of $H_L$ and $H_R$:
Using the resolution of the identity $\sum_{\bm l, \bm r} |\bm l \bm r\rangle \langle \bm l \bm r| = \mathbb 1$, the
final quantum entropy reads:
\begin{align}
S_1&= k_B\frac{\sum_{\bm n,\bm l, \bm r} P(\bm n | \bm l , \bm r) e^{-\beta_L E_{\bm l} }  e^{-\beta_R E_{\bm r} }  \ln \nu_{\bm n} } {Z(\beta_L)Z(\beta_R)}\, , 
\label{eq:S1}
\end{align}
where 
\begin{align}
P(\bm n | \bm l , \bm r) = \langle \bm n| \bm l \bm r\rangle \langle \bm l \bm r  |\bm n\rangle\, .
\end{align}
In order to calculate $P(\bm n | \bm l , \bm r)$ we consider the basis 
$|\bm s \rangle= |s_1, s_2 \dots , s_N \rangle = |s_1 \rangle \otimes |s_2 \rangle \otimes \dots \otimes|s_N \rangle$  formed by the direct 
product of the eigenstates $|s_j \rangle$ of the $z$ component of each spin operator: 
$\sigma^z_j | s_j \rangle =s_j |s_j \rangle$ with $s_j=\pm$.
Employing the resolution of the identity twice we obtain:
\begin{align}
P(\bm n | \bm l , \bm r)&= \sum_{\bm s,\bm s'} \langle \bm n| \bm s \rangle \langle \bm s| \bm l \bm r\rangle 
\langle \bm l \bm r| \bm s' \rangle \langle \bm s'  |\bm n\rangle \nonumber  \\
&= \sum_{\bm s,\bm s'} \langle \bm n | \bm s \rangle \langle \bm s'  |\bm n\rangle
\langle \bm l \bm r| \bm s' \rangle  \langle \bm s| \bm l \bm r\rangle \, .
\label{eq:Pnlr}
\end{align}
The next crucial step in the calculation consists in expressing the operators 
$ | \bm s \rangle \langle \bm s'  |$ in terms of spin rising and lowering operators.
For a single spin, say spin $j$, we have
$
|+ \rangle \langle + | = \sigma_j^+\sigma_j^- \, ,|- \rangle \langle - | = \sigma_j^-\sigma_j^+ \, ,
|+ \rangle \langle - | = \sigma_j^+ \, ,|- \rangle \langle + | = \sigma_j^-
$
which can be compactly written
\begin{align}
|s \rangle \langle s' |= \delta_{ss'} \sigma_j^s\sigma_j^{-s} + (1-\delta_{ss'})\sigma_j^s\, ,
\end{align}
where $\delta_{ss'}$ denotes the Kronecker symbol, and $\sigma_j^\pm$ are given in Eq. (\ref{eq:sigmas}).
Therefore:
\begin{align}
| \bm s \rangle \langle \bm s'  |= \prod_j[\delta_{s_j s_j'} \sigma_j^{s_j}\sigma_j^{-s_j} + (1-\delta_{s_js_j'})\sigma_j^{s_j}]\, .
\label{eq:lsXsl}
\end{align}
In order to calculate $\langle \bm n | \bm s \rangle \langle \bm s'  |\bm n\rangle$ we express 
$| \bm s \rangle \langle \bm s'  |$ in terms of the fermionic operators $b_k$ of the total system.
This can be accomplished by using
the inverse sine transforms:
\begin{align}
 a_j &= \sqrt{\frac{2}{N+1}} \sum_{k=1}^{N} \sin \left( \frac{kj \pi}{N+1}
\right)b_k\, ,
\label{eq:inverse}
\end{align}
and the following identities:
\begin{align}
\sigma_j^s \sigma_j^{-s} &= a_j^s a_j^{-s} \label{eq:sigmaj2} \, ,\\ 
\sigma_j^{-s} &= \prod_{k=1}^{j-1}(2 a_k^+a_k^- -1)a_j^{-s} \, ,\label{eq:sigmaj1}
\end{align}
where we set the convenient notations $a_j^+\doteq a_j^\dagger$, $a_j^-\doteq a_j$.
By plugging Eqs. (\ref{eq:inverse}, \ref{eq:sigmaj2}, \ref{eq:sigmaj1}) into Eq. (\ref{eq:lsXsl}), the operator
$| \bm s \rangle \langle \bm s'  |$ is expressed in terms of the operators $b_k$ and $b_k^\dagger$.
The wanted term $\langle \bm n | \bm s \rangle \langle \bm s'  |\bm n\rangle$ is calculated then
by using the properties (\ref{eq:bk1}-\ref{eq:bk3})
and the orthonormality condition $\langle \bm n| \bm n'\rangle = \delta_{n_1,n_1'}\delta_{n_2,n_2'}\dots \delta_{n_N,n_N'}$.

In order to calculate $\langle \bm l \bm r| \bm s' \rangle  \langle \bm s| \bm l \bm r\rangle $
we proceed by expressing $| \bm s \rangle$ as the direct product of two
subchain states 
$| \bm s \rangle=| s_1  \dots s_{N}\rangle = | s_1 \dots s_{N/2}\rangle \otimes  | s_{N/2+1} \dots s_{N}\rangle =
 | \bm s_L \rangle \otimes | \bm s_R \rangle$. 
Accordingly, the wanted term reduces to the product of two
terms each pertaining  to each subchain: 
$\langle \bm l \bm r| \bm s' \rangle  \langle \bm s| \bm l \bm r\rangle$ $ =
\langle \bm l | \bm s'_L \rangle  \langle \bm s_L | \bm l\rangle \cdot  \langle \bm r | \bm s'_R \rangle  \langle \bm s_R | \bm r\rangle  $.
The calculation of $\langle \bm l | \bm s'_L \rangle  \langle \bm s_L | \bm l\rangle$ proceeds as
the above calculation of $\langle \bm n | \bm s \rangle \langle \bm s'  |\bm n\rangle$, with the only difference
that now all quantities pertain to the smaller $L$-chain. Since the $L$ and $R$ chains are identical this automatically gives also the $R$-chain term $\langle \bm r | \bm s'_R \rangle  \langle \bm s_R | \bm r\rangle$.

In the Appendix we provide further details on how these calculations were 
implemented.

\section{Entropy change}
\label{sec:entropychange}
\begin{figure}
\includegraphics[width=0.48\textwidth]{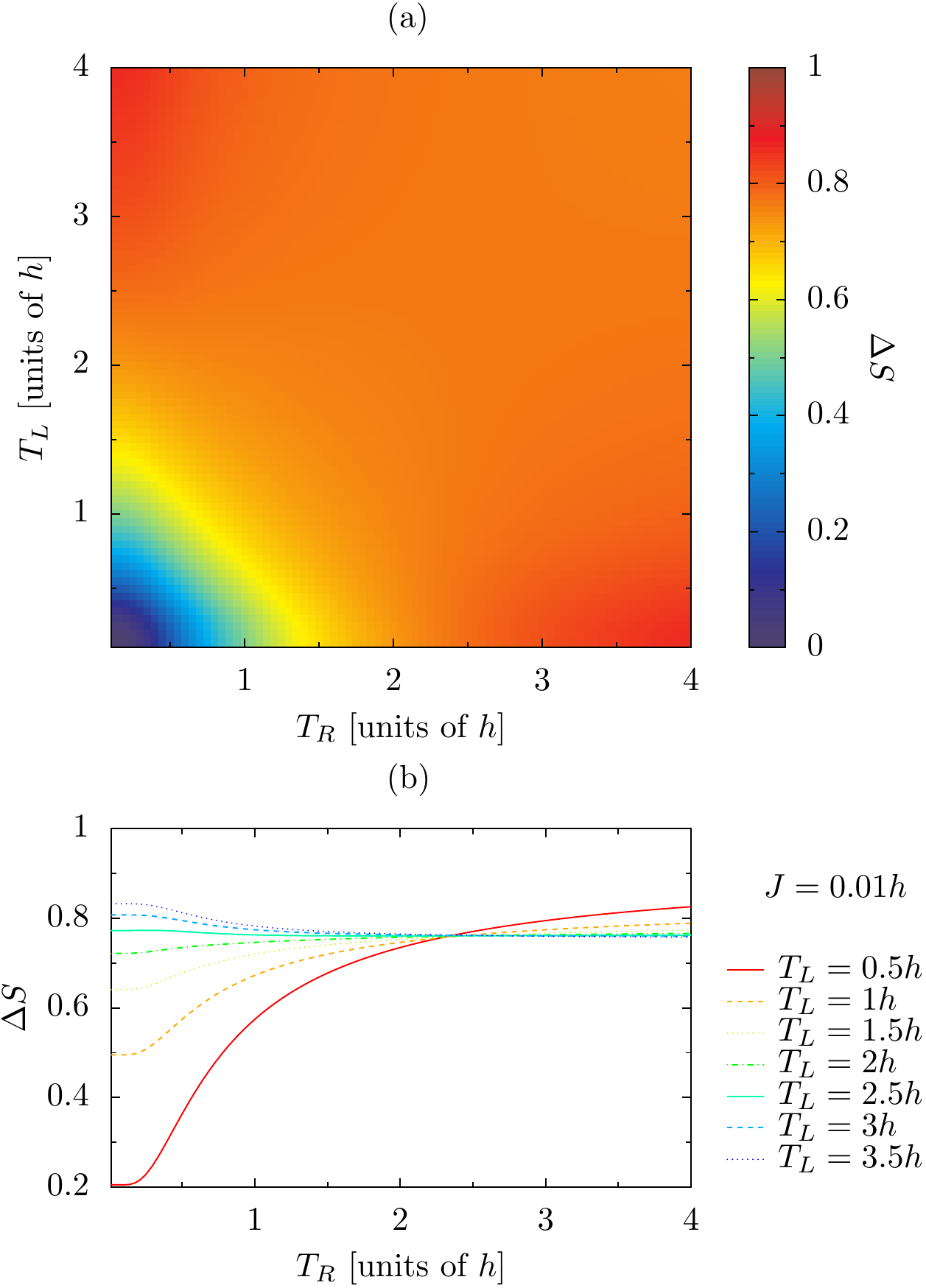}
	\caption{(a): Entropy change, Eq. (\ref{dS}), as a function of $T_L,T_R$. 
		     (b): Entropy change, Eq. (\ref{dS}), as a function of $T_R$ for various fixed $T_L$'s. Here $h=1$, $J=0.01h$ and $N=8$.}
	\label{fig:2}
\end{figure}
\begin{figure}[]
		{\includegraphics[width=0.48\textwidth]{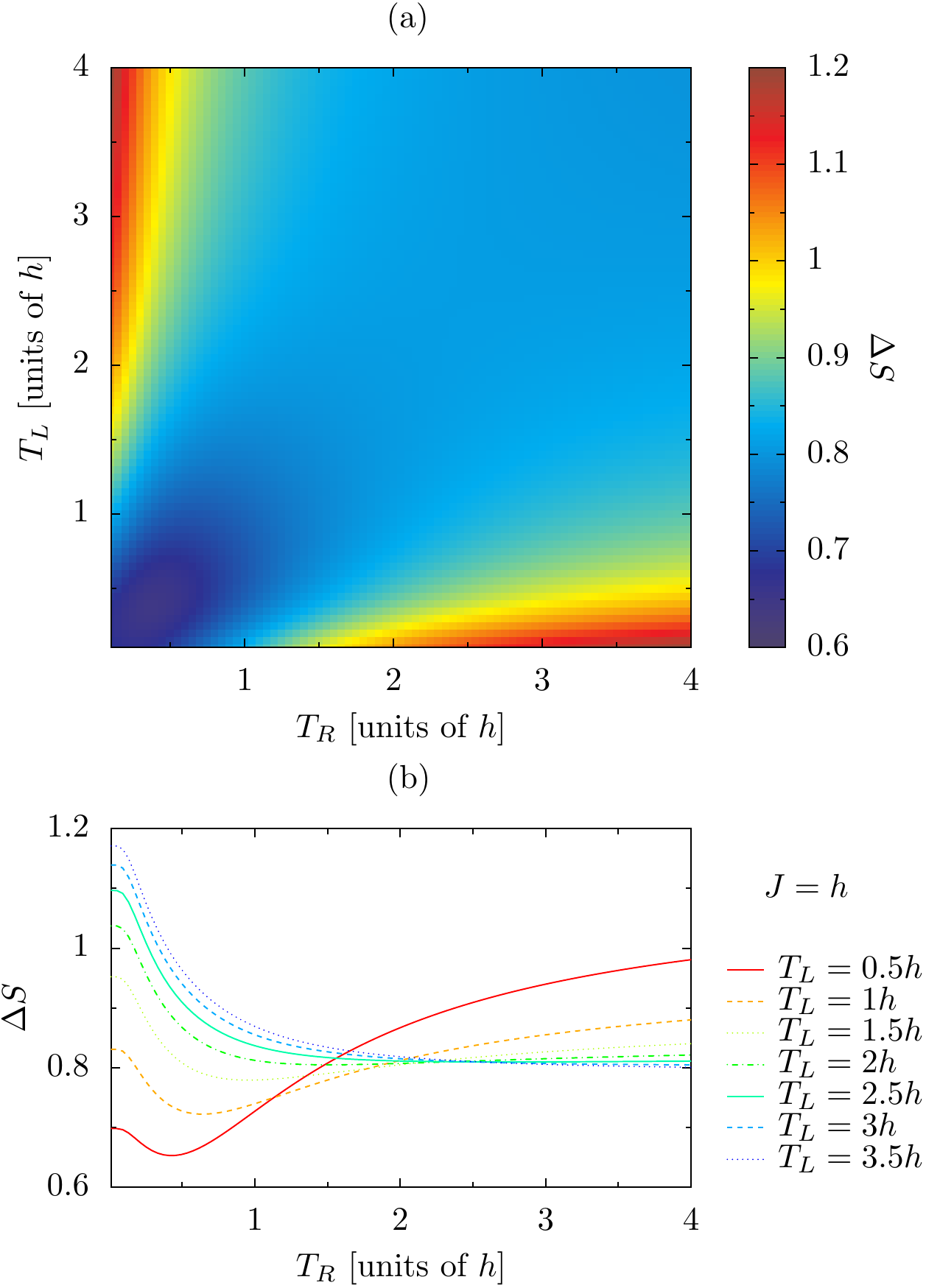}}
\caption{(a): Entropy change, Eq. (\ref{dS}), as a function of $T_L,T_R$. 
		     (b): Entropy change, Eq. (\ref{dS}), as a function of $T_R$ for various fixed $T_L$'s. Here $h=1$, $J=h$ and $N=8$.}
	\label{fig:3}
\end{figure}
Using the method detailed above, we have calculated the change of quantum entropy
\begin{align}
\label{dS}
\Delta S = S_{1} - S_{0}
\end{align}
caused by the sudden quench, for different values of the parameters defining 
our problem. These are the initial temperatures of the left and right chains, $k_B T_{L,R} = \beta_{L,R}^{-1}$,
the length $N$ of the total chain, and the interaction strength $J$. The field strength $h$ was used as the unit of energy, and we adopted the convention that temperatures are measured in units of energy. In these units we have $k_B$, Boltzmann's constant, equal to $1$.

Figure \ref{fig:2}, top panel, shows a surface plot of the entropy change in a chain of length $N=8$ at $J=0.01 h$,
as a function of $T_L$ and $T_R$.
Figure \ref{fig:2}, bottom panel, shows for the same values of $N=8, J=0.01h$,
the behavior of $\Delta S$ as function of $T_R$ for various left chain temperatures $T_L = 0.5h$ to $3.5 h$.
The quantum entropy change here is always positive and approaches a saturation value as $T_R$ becomes very large. 
We see qualitatively different behaviors of $\Delta S$ as a function of $T_R$, depending on $T_L$.
As $T_L$ increases, $\Delta S$ changes from a monotonically increasing function of $T_R$ to a monotonically 
decreasing function  of $T_R$. The transition occurs for $T_L$ of the order of the width of the spectrum of $H_L$ 
(which takes on the value $2.6h$ in this case).

In Fig. \ref{fig:3} we report the same quantities as in Fig. \ref{fig:2}
with the only difference that now $J= h$. For this value $J=h$ the quantum entropy change, seen as a function
of $T_R$ features minima for small values of $T_L$ and a monotonically decreasing behavior for large $T_L$.
As with Fig. \ref{fig:2} the transition occurs for $T_L$ of the
order of the width of the spectrum of $H_L$ (which is $2h$ in this case).

In both Fig.  \ref{fig:2} and \ref{fig:3}, we see a common feature. The values of $\Delta S$ corresponding to lower $T_L$ 
is smaller than the value of $\Delta S$ corresponding to larger $T_L$ at lower $T_R$, whereas it is vice-versa at larger values of $T_R$. 
We observed a similar behavior also at other values of $J$ (e.g., for  $J= 0.0001, 0.1, 2, 5$) and  $N=2,4,6$ for
same value of all other parameters). In no case have we observed a negative change in quantum entropy.

For $T_L=T_R$, $\Delta S$ gives the entropy of mixing. In the high temperature limit $T_L=T_R \gg \Delta E$, where $\Delta E$
is the width of the spectrum, one gets $\Delta S = 2^{-N} \ln( 2^N!)-2^{-N/2+1}\ln({2^{N/2}!})$. For $N=8$, as in Figs.
\ref{fig:2} and \ref{fig:3}, this gives $\Delta S  \simeq 0.72$. For large $N$, using Stirling approximation one gets
$\Delta S \simeq 1$. The mixing entropy is non-zero because the spin chain is made of distinguishable particle (one can distinguish one
spin from the other by its site label). It is however negligibly small as compared to the entropy itself, which is of order $N$, because it is a surface effect, and as such is of order 1.

\section{Thermalization}
\label{sec:thermalization}
After the quench, the system is in an out-of-equilibrium state. 
In order to quantify how far the system is from an equilibrium 
Gibbs state, one can employ one of the many metrics in the space of density matrices
discussed in the literature, e.g., in reference  \cite{Dajka11QIP10}.
Among them the Hilbert-Schmidt distance:
\begin{align}
D_{HS} [\rho,\sigma] = \sqrt{\Tr(\rho - \sigma)^2}
\label{eq:H_HS}
\end{align}
appears best suited to the problem at hand.
The reason is that, in our problem, the Hilbert-Schmidt distance between the after-quench density matrix $\rho_{t}$
and a Gibbs state
\begin{align}
\rho_\beta=e^{-\beta H_1}/Z_1(\beta)\, ,
\end{align}
where 
$Z_1(\beta)=\Tr\,  e^{-\beta H_1}$, does not 
depend on time $t$. Furthermore it can be calculated by knowing the initial
density matrix $\rho_0$, Eq. (\ref{eq:initialstate}), and the transition 
amplitudes $P(\bm{n}, \bm{lr})$, Eq. (\ref{eq:Pnlr}). That is, it can be obtained
from the only knowledge of the (time independent) diagonal elements
\begin{align}
\langle \bm{n}|\rho_t|\bm{n}\rangle = \sum_{\bm{lr}} P(\bm{n}|\bm{l},\bm{r}) \frac{e^{-\beta_L E_{\bm l} }  e^{-\beta_R E_{\bm r} }} {Z(\beta_L)Z(\beta_R)}\, , 
\end{align}
with no need to calculate the off-diagonal elements\\
 $\langle \bm{n}|\rho_t|\bm{m}\rangle$, $\bm{n}\neq\bm{m}$.
 In fact:
\begin{align}
D_{HS}^2 [\rho_t,\rho_\beta] =\Tr \rho_\beta^2 -2\Tr \rho_t \rho_\beta + \Tr \rho_t^2\, ,
\end{align}
and
\begin{align}
\Tr \rho_\beta^2 &= \sum_{\bm{n}} \frac{e^{-2\beta E_{\bm{n}}}}{Z_1^2(\beta)}\, , \\
\Tr \rho_t \rho_\beta &= \sum_{\bm{n}} \frac{e^{-\beta E_{\bm{n}} } } {Z_1(\beta)} \langle \bm{n}|\rho_t|\bm{n}\rangle\, ,\\
\Tr \rho_t^2 &= \Tr \rho_0^2= \sum_{\bm{rl}} \frac{e^{-2\beta_L E_{\bm{l}}}}{Z^2(\beta_L)} \frac{e^{-2\beta_R E_{\bm{r}}}}{Z^2(\beta_R)}\, .
\label{eq:Tr_rho_t_2}
\end{align}
Using Eqs. (\ref{eq:H_HS}-\ref{eq:Tr_rho_t_2}) we calculated the minimal distance
\begin{align}
D_{HS}^\text{min}= \min_\beta D_{HS} [\rho_t,\rho_\beta] \, ,
\label{eq:DHSmin}
\end{align}
between the final state $\rho_t$ and the set of thermal Gibbs states. This
gives both an estimate of how far the system is from equilibrium, and what 
the temperature $k_B \bar{T}={\bar{\beta}}^{-1}$ is of the closest equilibrium,
where $\bar{\beta}$ is the value of $\beta$ for which the minimum distance
$D_{HS}^\text{min}$ is attained.
\begin{figure}[h!]
\includegraphics[width=0.48\textwidth]{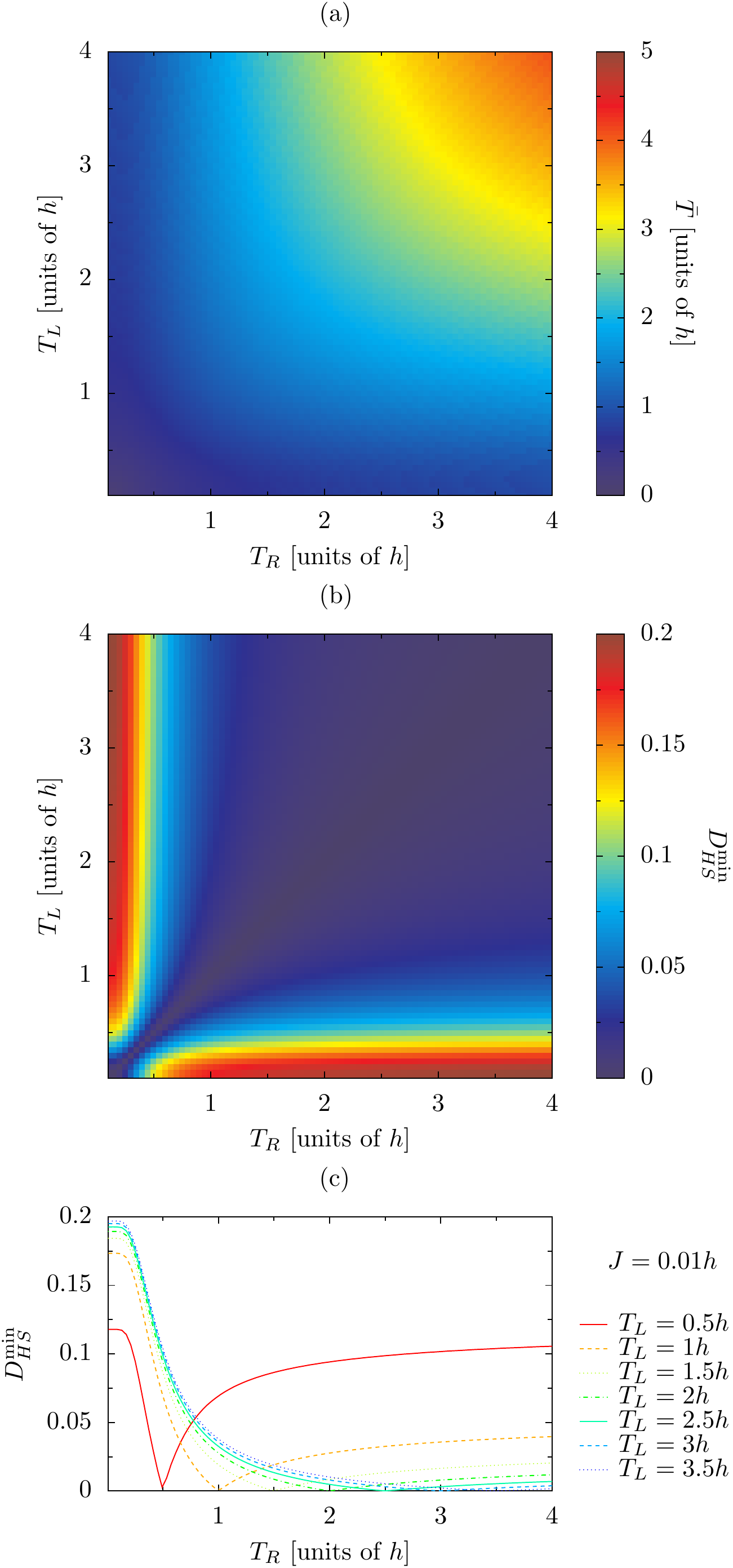}
\caption{
	(a): Temperature $\bar T$ of closest equilibrium, as a function of $T_L,T_R$. 
		     (b): Minimal distance $D_{HS}^\text{min}$, Eq. (\ref{eq:DHSmin}), as a function of $T_R,T_R$.	
		     (c): Minimal distance $D_{HS}^\text{min}$, Eq. (\ref{eq:DHSmin}), as a function of $T_R$ for various fixed $T_L$'s.
		     	Here $h=1$, $J=0.01h$ and $N=8$.}
	\label{fig:4}
\end{figure}
\begin{figure}[h!]
\includegraphics[width=0.48\textwidth]{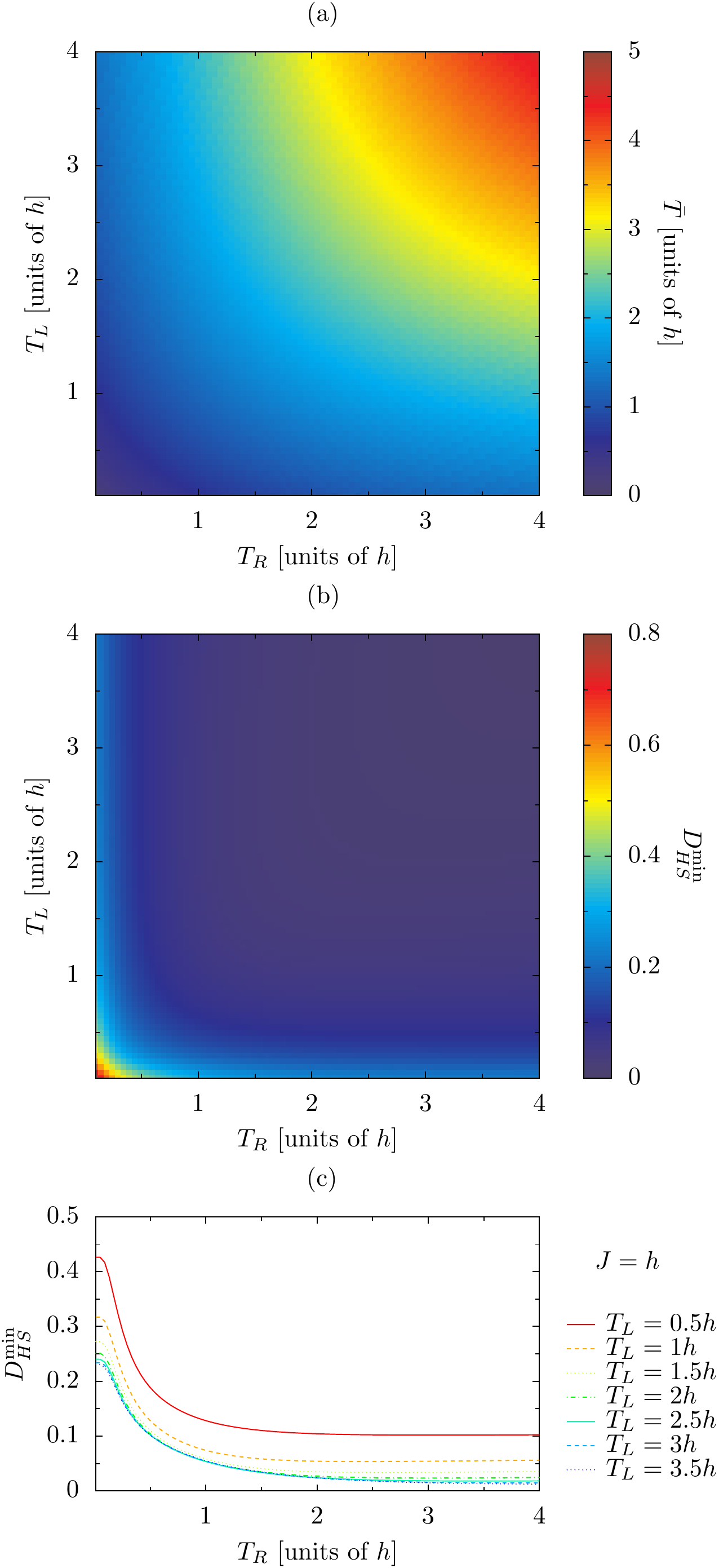}
\caption{	(a): Temperature $\bar T$ of closest equilibrium, as a function of $T_L,T_R$. 
		     (b): Minimal distance $D_{HS}^\text{min}$, Eq. (\ref{eq:DHSmin}), as a function of $T_R,T_R$.	
		     (c): Minimal distance $D_{HS}^\text{min}$, Eq. (\ref{eq:DHSmin}), as a function of $T_R$ for various fixed $T_L$'s.
		     	Here $h=1$, $J=h$ and $N=8$.}
\label{fig:5}
\end{figure}

Fig. \ref{fig:4} shows plots of $\bar{T}$ and $D_{HS}^\text{min}$ (panels (a) and (b), respectively),
as a function of $T_L,T_R$, at a low value of interaction strength $J=0.01 h$ and chain length $N=8$. 
Panel (c) presents $D^\text{min} _{HS}$ as function of $T_R$ for various fixed $T_L$'s.
Panel (b) indicates that better thermalization is achieved when $T_L$ and $T_R$ are closer.
As can be seen from panel (a), when $T_L=T_R$ it is $\bar{T} \simeq T_L=T_R$.
For fixed $T_L$ (panel (c)) we observe that as $T_R$ grows from zero, the minimal distance
$D^\text{min}_{HS}$ first decreases, then reaches a minimum, and finally grows. The minimum is 
in correspondence to $T_R \sim T_L$ as expected, and it is sharper for low $T_L$'s.

Fig. \ref{fig:5} is like Fig. \ref{fig:4} with the only difference that it is for a larger
interaction strength $J=h$.
As compared with Fig. \ref{fig:4}  we observe here a different structure of the plot of
$D^\text{min} _{HS}$ as a function of $T_L$ and $T_R$, see panel (b).
Within the range of $T_L$, $T_R$ considered in the plot, it appears that a smaller distance
is achieved when either $T_L$, $T_R$ or both grow.
This different structure is reflected in the curves presented in panel (c), presenting $D^\text{min} _{HS}$ as a function of 
$T_R$, for fixed $T_L$'s. The structure of the plot of $\bar{T}$ as a function of
$T_L,T_R$, panel (a), is qualitatively similar to the corresponding plot in Fig. \ref{fig:4}.
Note that for $T_L=T_R$, the temperature of the closest equilibrium is larger than $T_L$,
of some amount of the order of $h$. This is because the quench injects an energy amount
of the order $J=h$ in the system.
Similarly, in Fig. \ref{fig:4}, $\bar{T}$ was close to $T_L=T_R$ because the 
quench injects in that case an energy of the order $J=0.01h $.
\section{Conclusions}\label{sec:conclusions}
We have investigated numerically the change in the quantum Hertz entropy of
Eq. (\ref{eq:hatS}), caused by a quench of two spin-chains of different temperatures into
a larger single chain.
Although we cannot conclude that such changes are always positive,
our numerics clearly suggests that this is the typical behavior, thus providing further support
to the statement that
the quantum Hertz entropy of Eq. (\ref{eq:hatS}) is a proper quantum entropy for thermally isolated systems. 

We further quantified how far the
system is from equilibrium after the quench, and estimated the corresponding temperature
of the closest equilibrium. For those quenches ending very close to an equilibrium state,
it becomes meaningful to assign the system the equilibrium temperature $\bar{T}$, and
the thermodynamic entropy $S_1$.

The Hertz entropy $S$ can be employed to study phase transitions and critical points in spin chains
in a way analogous to Ref.  \cite{Dorner12PRL109} where the dissipated work $W_\text{diss}$ 
signaled the crossing of a critical point as the magnetic field was incrementally and globally changed,
and the chain was initially at some temperature $T$. Because of the strict connection between Hertz entropy 
and dissipated work, the Hertz entropy in that same scenario should give similar results.
The present thermalization scenario, with an initial nonequilibrium state and a local quench,
is not convenient though, for the study of critical points.

\subsection*{Acknowledgements}
This work was supported by the DAAD-WISE 2011 Scholarship (D.G.J), the German-Israeli Foundation via grant no. G 1035-36.14/2009 (D.G.J) 
and the German Excellence Initiative Nanosystem Initiative Munich (M.C).
D.G.J expresses sincere thanks to IISER, Pune, India for providing this great opportunity.
D.G.J. would like to thank Prof. P. H\"anggi (University of Augsburg) for the kind hospitality
in his group. D.G.J. is also thankful to his friend Shadab Alam for fruitful discussions related to numeric. 

\appendix
\section*{Appendix}
As detailed in Sec. \ref{sec:FinalEntropy}, the transition
probabilities $P(\bm n | \bm l , \bm r)$, involve the calculation of
the expectation of the operators $ | \bm s \rangle \langle \bm s'  |$ over the Fock states
$| \bm{n} \rangle$.
Accordingly, we have detailed how these operators may be expressed
in terms of the fermonic operators $b_k$, whose action on the Fock states is defined
in Eqs. (\ref{eq:bk1},\ref{eq:bk2},\ref{eq:bk3}). In order to calculate those expectations we expressed
the fermionic operators $b_k$ in matrix form. First we represented the Fock states $| \bm n \rangle$
as tensorial product of single-spin states:
\begin{align}
| 0 \rangle = 
\left(
\begin{array}{clrr} 
0 \\ 
1 
\end{array}
\right)	\, , \qquad
| 1 \rangle = 
\left(
\begin{array}{clrr} 
1 \\ 
0 
\end{array}
\right)	\, .
\end{align}
For example, the Fock state $|01\rangle$ of a chain of $N=2$ spins read
\begin{align}
| 0 1 \rangle &= 
\left(
\begin{array}{clrr} 
0 \\ 
1 
\end{array}
\right)	\otimes
\left(
\begin{array}{clrr} 
1 \\ 
0 
\end{array}
\right) = 
\left(
\begin{array}{clrr} 
0 \\ 
0 \\
1	\\
0
\end{array}
\right)
\end{align}
and similarly for larger chains.
In this representation, the searched fermionic operators are represented by the following matrix tensorial products:
\begin{align}
b_k &= (-1)^{k-1}\underbrace{\sigma^z \otimes \ldots \otimes \sigma^z}_{\mbox{$k-1$ terms}}\,  \otimes \, {\sigma^-} \otimes 
\, \underbrace{\mathbb 1 \otimes \ldots \otimes\,  \mathbb 1}_{\mbox{$N-k$ terms}}	\\
b_k &= (-1)^{k-1}\underbrace{\sigma^z \otimes \ldots \otimes \sigma^z}_{\mbox{$k-1$ terms}}\,  \otimes \, {\sigma^+} \otimes 
\, \underbrace{\mathbb 1 \otimes \ldots \otimes\,  \mathbb 1}_{\mbox{$N-k$ terms}}	\\
\end{align}
where $\sigma^{\pm}= (\sigma^x\pm \sigma^y)/2$ are rising and lowering operators, expressed in terms of the  Pauli matrices  $\sigma^{x,y,z}$, and $\mathbb 1$ is the $2 \times 2$ identity matrix.

The calculation of $P(\bm n | \bm l , \bm r)$ further requires the calculation of the
expectations $\langle \bm{l} | \bm s_L \rangle \langle \bm s'_L  |\bm{l}\rangle$,
$\langle \bm{r} | \bm s_R \rangle \langle \bm s'_R  |\bm{r}\rangle$. The calculation
of these proceeds exactly in the same way detailed above, with the only difference
that the chain length is now $N/2$ instead of $N$. With all these expectations one can calculate the probabilities $P(\bm n | \bm l , \bm r)$,
and, in turn, via Eq. (\ref{eq:S1}), the final quantum entropy.

The performance of the calculation can be greatly improved if one notices
the following selection rules
\begin{align}
&2\sum_{k=1} ^{N} {n_k} \neq \sum_{j=1} ^{N} {s_j} +N \implies \langle \bm{n} | \bm{s} \rangle = 0\, , \\
&2\sum_{k=1} ^{N/2} {l_k} \neq \sum_{j=1} ^{N/2} {s_j} +N/2  \implies \langle \bm{l} | \bm{s}_L \rangle = 0\, , \\
&2\sum_{k=1} ^{N/2} {r_k} \neq \sum_{j=N/2+1} ^{N} {s_j} +N/2  \implies \langle \bm{r} | \bm{s}_R \rangle = 0\, .
\end{align}
Together with Eq. (\ref{eq:Pnlr}) these rules imply that the quench at time $t=0$ conserves the number of excitations.

\end{document}